\documentclass[sigconf]{acmart}
\AtBeginDocument{%
  }

\copyrightyear{2026}
\acmYear{2026}
\setcopyright{cc}
\setcctype{by}
\acmConference[VRST '26]{32nd ACM Symposium on Virtual Reality Software and Technology}{November 16--18, 2026}{Sendai, Japan}
\acmBooktitle{32nd ACM Symposium on Virtual Reality Software and Technology (VRST '26), November 16--18, 2026, Sendai, Japan}
\acmDOI{10.1145/3822517.3848660}
\acmISBN{979-8-4007-2811-2/2026/11}

\begin{document}


\title{Gaze Under Strain: How Oculomotor Behavior Tracks Cybersickness Progression in VR}

\author{Nevzat Umut Demirseren}
\affiliation{%
\department{Department of Computer Science}
\institution{Texas State University}
\city{San Marcos}
\state{Texas}
\country{USA}}
\email{ndemirseren@txstate.edu}

\author{Mehedi Hasan Raju}
\affiliation{%
\department{Department of Computer Science}
\institution{Texas State University}
\city{San Marcos}
\state{Texas}
\country{USA}}
\email{m.raju@txstate.edu}

\author{Oleg Komogortsev}
\affiliation{%
\department{Department of Computer Science}
\institution{Texas State University}
\city{San Marcos}
\state{Texas}
\country{USA}}
\email{ok11@txstate.edu}

\author{Isayas Berhe Adhanom}
\affiliation{%
\department{Department of Computer Science}
\institution{Texas State University}
\city{San Marcos}
\state{Texas}
\country{USA}}
\email{isayas@txstate.edu}

\begin{abstract}
Cybersickness remains a major barrier to comfortable, accessible virtual reality (VR). Arising from mismatches between visual and vestibular signals, it places sustained demands on gaze stabilization and visual processing. Because gaze is the mechanism through which users actively sample the virtual environment, cybersickness-induced disruptions may alter both what users perceive and how discomfort itself emerges. Yet, how eye-movement behavior evolves alongside discomfort during prolonged VR exposure remains poorly understood.
We present a behavioral analysis of eye-tracking data from 19 participants navigating a realistic virtual environment over a 20-minute session, with discomfort rated repeatedly throughout the experiment. Linear mixed-effects models examined gaze constructs spanning fixation stability, visual exploration, and gaze allocation as functions of discomfort, exposure duration, and their interaction. Our results show that fixation stability degraded progressively with increasing discomfort, with the effect strengthening over time. Visual exploration showed no marginal discomfort effect but exhibited significant discomfort × time interactions, indicating that saccadic behavior adapts dynamically rather than responding uniformly to discomfort. Gaze allocation was not associated with discomfort but shifted systematically with time. These findings show that cybersickness does not uniformly disrupt gaze. Instead, different aspects of eye-movement behavior respond differently and change in ways that only become visible when discomfort is tracked repeatedly over time. Understanding  how gaze evolves during cybersickness progression provides the behavioral grounding needed to better interpret how discomfort shapes the user's visual experience of immersive environments and to design more sensitive detection approaches.

\end{abstract}



\maketitle

\section{Introduction}

Cybersickness continues to limit who can comfortably use virtual reality (VR) and for how long, with symptom severity and susceptibility varying widely across users \cite{laviola20173d, rebenitsch2016review, stanney2020identifying}. This variability is both an accessibility concern and a measurement problem. Despite advances in displays, tracking, and rendering, cybersickness remains a persistent challenge, and the field still lacks fine-grained, behaviorally grounded accounts of how it develops. Cybersickness is most often quantified through subjective self-report \cite{kennedy1993simulator, keshavarz2011validating}, which captures whether and how much discomfort occurred but reveals little about the perceptual and oculomotor processes that accompany its onset and growth.

A growing body of work has linked cybersickness to physiological and behavioral signals, including heart rate variability, electrodermal activity, postural sway, head motion, and eye tracking \cite{cortes2023eeg, jeong2019cybersickness, li2023exploring, tian2024says, dennison2016use, nalivaiko2015cybersickness, garcia2019identifying, jeong2022eyes, islam_cybersickness_2021}. These studies establish that user behavior changes as discomfort develops and have driven substantial progress in detection and prediction. Their primary aim, however, is typically to forecast a sickness label rather than to describe how the underlying behavior unfolds, so the temporal dynamics of discomfort progression remain comparatively underexplored.

Eye movements are a particularly informative window onto these dynamics. Cybersickness is closely tied to visual motion processing and oculomotor control, and gaze is the mechanism through which users actively sample a virtual scene; disruptions to gaze therefore bear directly on both what users perceive and how discomfort emerges. Prior eye-tracking research supports this link, relating fixation and saccade measures to discomfort and showing that gaze supports user-state modeling \cite{nam_eye_2022, ozkan_relationship_2023, islam_cybersickness_2021, kundu_litevr_2023}. Much of this evidence, however, comes from predictive models, session-aggregated gaze features, condition-level contrasts, or post-exposure sickness scores. Each is well suited to its purpose, but none is designed to track how oculomotor characteristics co-evolve with discomfort \emph{as it accumulates}. As a result, a basic behavioral question is left open: when discomfort is measured repeatedly throughout exposure, how does eye gaze behavior change in step with it?

We address this question through a behavioral analysis of eye-tracking data from a prolonged, realistic VR navigation task in which discomfort was rated repeatedly across a 20-minute session. Rather than using gaze to predict a sickness label, we ask how interpretable fixation- and saccade-derived metrics co-evolve with discomfort over time. Reusing this time-resolved dataset lets us model not only the marginal relationship between discomfort and gaze but also how that relationship changes with continued exposure, an effect that aggregate or post-hoc designs cannot resolve. Our analysis reveals that this temporal dimension is essential: fixation stability degrades as discomfort rises and does so increasingly over exposure, and gaze allocation shifts with exposure duration independently of discomfort. Cybersickness, in short, does not disrupt gaze uniformly, and several of these patterns are visible only when discomfort is tracked over time.

This paper makes the following contributions:
\begin{itemize}
    \item A time-resolved behavioral characterization of how fixation and saccade metrics relate to discomfort during prolonged VR exposure, organized around fixation stability, visual exploration, and gaze allocation.
    \item Evidence that the discomfort-gaze relationship is \emph{time-varying}: fixation duration shows a strengthening negative association with discomfort, and saccadic exploration shows significant discomfort-by-time interactions, demonstrating effects that aggregate or post-exposure measures cannot capture.
    \item A behavioral interpretation that complements existing predictive and aggregate-feature approaches, providing grounding for understanding how discomfort shapes the user's visual experience and for more sensitive, temporally aware cybersickness detection.
\end{itemize}

\section{Background}\label{sec:background}

\subsection{Causes of Cybersickness}

Although no single theory fully explains cybersickness \cite{stanney2020identifying, tian2022review}, several established frameworks provide insight into its underlying mechanisms.

\subsubsection{Sensory Conflict Theory}

Sensory Conflict Theory is the most widely accepted explanation of cybersickness \cite{reason1975motion, oman_motion_1990}. It proposes that discomfort arises when visual motion cues conflict with vestibular and proprioceptive signals. In VR, users often perceive self-motion visually without corresponding physical motion, producing intersensory conflict that may contribute to symptoms such as nausea, dizziness, and disorientation.

\subsubsection{Postural Instability Theory}

Postural Instability Theory argues that cybersickness results from prolonged difficulty maintaining postural control rather than sensory conflict alone \cite{riccio1991ecological, stoffregen1998postural}. According to this perspective, instability in balance regulation precedes sickness onset, and symptoms persist until stable control is reestablished.

\subsubsection{Eye Movement Theory}

Eye Movement Theory proposes that motion sickness symptoms may arise from excessive oculomotor strain during visually induced eye movements \cite{ebenholtz_motion_1992, ebenholtz_possible_1994}. Prolonged exposure to visual motion may increase demands on gaze stabilization and eye movement control, suggesting that changes in fixation and saccadic behavior may accompany the development of cybersickness.

\subsection{Measurement of Cybersickness}

\subsubsection{Subjective Measures of Cybersickness}

Cybersickness is commonly assessed using subjective self-report instruments. The Simulator Sickness Questionnaire (SSQ) provides a standardized post-exposure assessment across nausea, oculomotor discomfort, and disorientation symptoms \cite{kennedy1993simulator}. Alternative measures, such as the Fast Motion Sickness (FMS) scale and discomfort score questionnaires, support repeated assessments during VR exposure and enable finer-grained tracking of symptom progression over time \cite{keshavarz2011validating, fernandes2016combating, adhanom2020effect}. Despite their widespread use, subjective measures rely on self-report, may interrupt immersion, and provide limited insight into the real-time development of cybersickness symptoms.

\subsubsection{Objective Measures of Cybersickness}

To complement subjective assessments, prior research has investigated physiological and behavioral indicators of cybersickness, including electrodermal activity (EDA), postural sway, body temperature, blink rate, and heart rate \cite{magaki2017measuring, hakkinen2002postural, gervsak2020effect, lopes2020exploring, yamamura2020pleasant}. Behavioral measures derived from VR tracking data, particularly postural behavior, have also been associated with cybersickness susceptibility \cite{arcioni2019postural, palmisano2018predicting, riccio1991ecological, risi2019effects, stoffregen1998postural, deguzman2025reduction}. 
While these approaches provide objective correlates of discomfort, many focus on predicting cybersickness using multimodal signals rather than characterizing how underlying perceptual and behavioral processes evolve during immersive interaction.

\subsection{Eye Movements and Cybersickness}

Eye movements provide a direct window into visual processing and oculomotor control, both of which are closely implicated in cybersickness. Because immersive VR environments often induce visually driven self-motion, users must continuously stabilize gaze, coordinate eye movements, and allocate visual attention under conditions of sensory conflict. As a result, gaze behavior offers a natural and behaviorally grounded means of examining how users interact with dynamic virtual environments as discomfort develops.

Prior work has shown that eye-tracking signals are sensitive to changes in user state during VR exposure. Variations in fixation duration, saccade amplitude, blink behavior, and pupil dynamics have been associated with discomfort and visually induced motion \cite{nam_eye_2022, ozkan_relationship_2023}. More broadly, research in related domains has demonstrated that eye-movement behavior reflects underlying perceptual and cognitive processes, including attentional allocation, cognitive load, and perceptual stability \cite{foucher2025eye, wang2025microsaccades}. These findings suggest that gaze behavior may provide insight into how visual processing is affected when sensory integration is disrupted, as in cybersickness.

Despite this potential, much of the existing VR literature has primarily leveraged eye tracking as a signal for modeling or predicting user discomfort, often in combination with other physiological measures \cite{dennison2016use, lopes_eye_2020, islam_cybersickness_2021, kundu_litevr_2023, chang_predicting_2021}. Other studies have investigated associations between specific gaze metrics and cybersickness-related discomfort \cite{nam_eye_2022, ozkan_relationship_2023}.

Most recently, Tovar et al. \cite{tovar_identifying_2024} induced visual discomfort through near-eye optical distortion during prolonged, naturalistic VR gameplay and applied representational similarity analysis to identify the head- and gaze-behavior markers most correlated with self-reported discomfort, including a temporal analysis in which this correspondence strengthened over exposure; the identified signal was dominated by blink and vestibulo-ocular behavior. Collectively, these works demonstrate that gaze behavior contains meaningful information related to cybersickness and visual discomfort. However, prior studies have largely emphasized prediction, condition-based comparisons, or associations measured at coarser temporal resolutions, leaving fine-grained temporal relationships between gaze behavior and evolving discomfort comparatively underexplored.

Analysis of gaze metrics derived from eye-movement events offers a complementary and more behaviorally interpretable perspective by focusing on discrete oculomotor events such as fixations and saccades \cite{Komogortsev2013SmoothPursuit}. These events represent fundamental units of visual behavior and provide insight into how users sample and process visual information during immersive interaction. However, comparatively less attention has been given to examining how event-derived gaze metrics systematically evolve alongside repeated measures of discomfort throughout the temporal progression of cybersickness during prolonged VR exposure. Characterizing these temporal gaze dynamics may provide insight into how visual exploration and oculomotor behavior adapt as discomfort develops over time.

\subsection{Temporal Dynamics of Gaze Behavior and Cybersickness}

Cybersickness develops progressively during VR exposure rather than appearing instantaneously, suggesting that the visual and oculomotor processes associated with discomfort may also evolve over time. Consistent with this view, prior research has shown that physiological and behavioral signals, including heart rate, EDA, and postural sway, exhibit temporal changes during VR exposure that reflect cybersickness progression \cite{nalivaiko2015cybersickness, dennison2016use, dennison2017cybersickness}.

In contrast, the temporal evolution of gaze behavior during cybersickness remains understudied. Existing studies often rely on session-level summaries or use gaze signals primarily for discomfort prediction \cite{islam2022towards, ridwan2025beyond, jeong2024precyse}. While these approaches demonstrate that gaze contains meaningful information related to cybersickness, they provide limited insight into how fixation- and saccade-derived behaviors change alongside repeated measures of discomfort throughout prolonged VR exposure.

Characterizing these temporal gaze dynamics may provide a more interpretable understanding of how visual exploration, attentional allocation, and oculomotor behavior adapt as cybersickness develops. Examining how event-derived gaze metrics evolve across repeated within-session discomfort ratings offers an opportunity to investigate the behavioral processes that accompany discomfort progression and may provide insight into the mechanisms through which cybersickness emerges during immersive interaction.

\section{Methodology}
\label{sec:methodology}
As discussed in Section~\ref{sec:background}, cybersickness may affect fixation stability, visual exploration, and gaze allocation. To examine these behaviors, we derive a set of gaze metrics from the recorded eye-tracking data, aggregate them within discomfort rating intervals, and relate them to repeated discomfort ratings over the course of VR exposure. This section describes the gaze processing pipeline used to detect fixations and saccades (Section~\ref{sec:pipeline}) and the metrics derived from these events (Section~\ref{sec:metrics}).

\subsection{Eye-Tracking Data Recording}

Eye-tracking data were acquired using the integrated binocular eye tracker in an HTC Vive Pro Eye head-mounted display (HMD) at a sampling rate of 120~Hz. The HMD rendered stimuli at a combined resolution of 2880$\times$1600 pixels across both eyes. Raw data were collected via the SRanipal C++ SDK \cite{sranipal_docs} and synchronized with the remaining cybersickness measurements described in Section~\ref{sec:measures} using Lab Streaming Layer (LSL)~\cite{kothe2014lab}.

The SRanipal SDK reported gaze data as binocular 3D vectors expressed in a head-centered coordinate frame, with each sample comprising a ray origin $(x, y, z)$ and a unit-direction vector $(x, y, z)$ relative to the HMD's local coordinate system. Because gaze direction is defined relative to the head rather than to world space, the signal
captures the orientation of the eyes within the head independently of HMD pose in the world. This property is consequential for the present analysis: event metrics such as fixation duration, dispersion, saccade amplitude, and gaze eccentricity all reflect eye-in-head movements and are therefore not confounded by rigid-body head rotations or translational HMD movement.   

\subsection{Eye Gaze Data Processing Pipeline}\label{sec:pipeline}

Eye-tracking data were processed using a multi-stage pipeline to ensure robust event classification and alignment with subjective discomfort ratings. Gaze direction data from the left eye were used for all analyses, as monocular signals are well suited for event classification and reduce sensitivity to asymmetric tracking loss or calibration drift common in head-mounted systems \cite{hooge2019gaze}. The left eye was selected consistently across all participants to ensure a uniform input to the processing pipeline.

Samples flagged as invalid by the eye tracker were first removed from the signal. Invalid samples corresponded to frames flagged by the SRanipal eye-tracking SDK as lacking reliable gaze estimation, typically due to temporary tracking loss, blink-related occlusion, or insufficient pupil detection confidence. Short gaps introduced by invalid samples were interpolated when their duration was less than 75~ms to preserve temporal continuity of the gaze signal, consistent with previous work \cite{liu2018elaborate, kleberg2025introducing}. The resulting signal was subsequently smoothed using a moving average filter, consistent with the noise-reduction procedures used in previous work~\cite{barone2023potential, kleberg2021restricted, TobiiIVTFilter}, with a window size of five samples to reduce high-frequency noise while maintaining the underlying motion structure. The moving average acts as a low-pass filter that reduces rapid sample-to-sample fluctuations caused by the measurement noise, which could otherwise introduce artificial changes in the subsequent angular velocity calculation and affect velocity-based event classification.

Inter-sample angular displacement was then computed between consecutive 3D gaze direction vectors, and angular velocity was calculated by dividing this displacement by the corresponding time difference between samples (approximately 8.33~ms at the native 120~Hz sampling rate).

Eye-movement events were identified using an offline implementation of the velocity-threshold (I-VT) algorithm \cite{salvucci2000identifying}, following established guidelines for parameter selection \cite{holmqvist2011eye, TobiiIVTFilter}.
Samples with angular velocity exceeding a threshold of $35^\circ$/s were classified as saccades, while lower-velocity segments were considered candidate fixations. Consecutive fixation segments were merged when separated by less than 75~ms and when their angular separation was below $0.5^\circ$ to account for brief interruptions in otherwise continuous fixations \cite{holmqvist2011eye, TobiiIVTFilter}.

To ensure robust event classification, minimum and maximum duration thresholds were applied. A minimum fixation duration of 60~ms and a maximum duration of 2000~ms were enforced, and segments outside these ranges were discarded. Saccades were required to have a minimum duration of 10~ms, a maximum duration of 150~ms, and a minimum amplitude of $1^\circ$. These parameters were selected to provide robustness against measurement noise while preserving sensitivity to rapid eye-movement dynamics at the 120~Hz sampling rate \cite{salvucci2000identifying, holmqvist2011eye, TobiiIVTFilter}.

\subsection{Eye Gaze Metrics}\label{sec:metrics}
From the detected eye-movement events, we derived five complementary metrics, organized into three groups that mirror our hypotheses: fixation stability, visual exploration, and gaze allocation. Rather than exhaustively computing all available eye-tracking features, we selected metrics that are theoretically tied to oculomotor processes implicated in cybersickness and that admit clear behavioral interpretation \cite{nam_eye_2022, ozkan_relationship_2023, foucher2025eye}. We did not analyze event counts, event-rate distributions, or event-ordering measures. These capture the occurrence and temporal organization of individual events, whereas our questions concern how the magnitude of fixation- and saccade-level metrics changes with discomfort over exposure. Such event-rate and sequence measures are a complementary direction for future work.

\subsubsection{Fixation Stability}

Fixations index how well users maintain stable visual anchoring, a process expected to be sensitive to perceptual strain and sensory conflict. We characterize fixation stability through two complementary measures. \emph{Fixation duration} captures \emph{temporal stability} by quantifying how long gaze remains settled on a location, whereas \emph{fixation dispersion} captures \emph{spatial stability} by measuring the angular spread of gaze samples within a fixation. Together, these measures distinguish whether discomfort primarily affects the persistence of gaze, its spatial precision, or both.

\subsubsection{Visual Exploration}
Saccades index how users sample the scene through rapid gaze shifts, reflecting visual exploration and oculomotor control. We characterize this through \emph{saccade amplitude}, the spatial extent of gaze shifts, and \emph{saccade velocity}, the speed of gaze transitions. Together they separate \emph{how far} users redirect gaze from \emph{how quickly} they do so, allowing changes in exploration breadth and exploration dynamics to be examined independently.

\subsubsection{Gaze Allocation}
We additionally computed \emph{gaze eccentricity} to quantify center-biased viewing. This metric is motivated by evidence that cybersickness in 360$^\circ$ video can be accompanied by increasingly center-focused gaze allocation \cite{lee_towards_2025}; we include it to test whether comparable center bias arises during interactive locomotion. Eccentricity was defined as the angular distance between the gaze direction vector $\mathbf{v}_{\text{gaze}}$ and a forward-facing reference vector $\mathbf{v}_{\text{center}}$:
\begin{equation}
\text{eccentricity} = \arccos\!\left( \mathbf{v}_{\text{gaze}} \cdot \mathbf{v}_{\text{center}} \right) \times \frac{180}{\pi}
\end{equation}
Lower values indicate gaze concentrated near the center of the visual field, and higher values gaze directed toward the periphery. Because gaze vectors from head-mounted eye trackers are expressed in a head-centered coordinate frame, eccentricity reflects gaze deviation relative to the head rather than in absolute world coordinates.

These metrics were segmented into non-overlapping 60-second windows aligned with the timing of discomfort score reporting intervals. Within each window, event-derived gaze metrics were aggregated using their mean values, producing window-level feature representations for subsequent linear mixed-effects modeling.
This aggregation matched the temporal granularity of discomfort reporting while providing interpretable summaries of oculomotor behavior.

\section{Dataset}
We base our analysis on eye-tracking data from a previously published VR locomotion experiment \cite{adhanom2022vr}.
This section summarizes only the methodological details relevant to the present analysis. Full experimental procedures, apparatus, and the broader study design are reported in the original paper~\cite{adhanom2022vr}.

We selected this dataset because it has the specific properties our research question requires: prolonged single-session exposure (up to 20 minutes), discomfort ratings collected repeatedly at one-minute intervals throughout exposure, and continuous eye tracking recorded during naturalistic,
free-viewing locomotion. Together, these enable a time-resolved analysis of how event-derived gaze metrics relate to discomfort and how these relationships change over exposure, which session-level or post-exposure datasets cannot support.

\subsection{Experiment Design}

The original experiment involved a prolonged VR locomotion task in which participants navigated a virtual environment by following visual waypoints using controller-based locomotion and head-steered navigation. Sessions lasted up to 20 minutes, enabling repeated discomfort measurements throughout immersive exposure.

The original dataset comprised two sub-experiments spanning two environments, one realistic and one abstract. We restricted the present analysis, a priori, to the realistic environment from the first session, in which participants performed a single continuous navigation task throughout the exposure. The realistic environment affords the naturalistic, unconstrained visual exploration most representative of everyday immersive use, while limiting the analysis to the first session avoids cross-session adaptation and habituation effects, yielding a cleaner view of discomfort progression within a single continuous exposure.


\subsection{Procedure}

At the first session, participants read an information sheet, provided informed consent, and completed a demographic questionnaire before receiving instructions on the navigation task and controller usage. A baseline SSQ was administered prior to VR exposure. Participants stood in place and were free to rotate their heads but did not physically walk; forward and lateral movement was controlled with an
Xbox~360 controller, and steering was achieved through head rotation. They navigated along predefined waypoints to the end of the environment, for approximately 20 minutes of total immersion. The environment imposed no stimulus-locked tracking task, allowing free gaze behavior throughout. At one-minute intervals throughout exposure, participants rated their discomfort on a 0--10 scale using a visual slider that disappeared immediately after a response was entered, minimizing disruption to natural visual exploration. After the navigation task, participants completed a post-exposure SSQ.

\begin{figure}[h]
  \centering
  \includegraphics[width=1.00\linewidth]{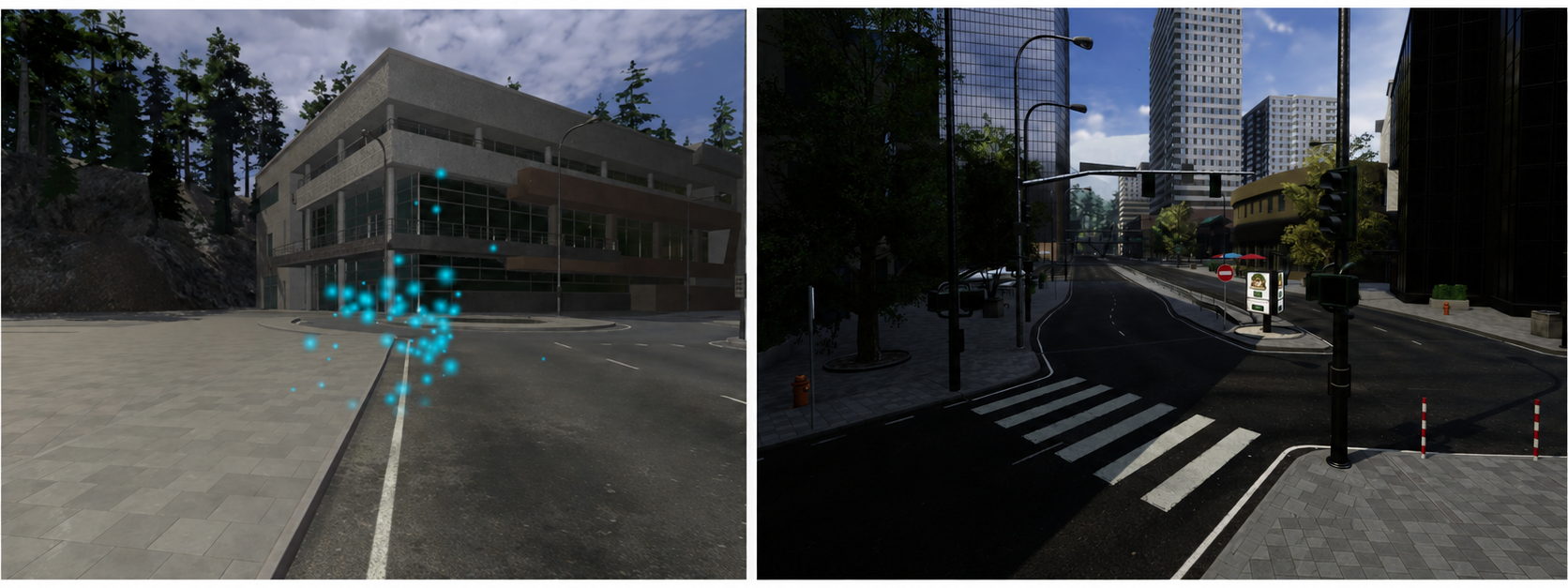}
  \caption{Representative screenshots from the experiment's virtual environment used for the study.}
  \Description{Images from the realistic virtual environment used for the study}
  \label{fig:environment}
\end{figure}

\subsection{Participants} 
A total of 22 participants (11 male, 11 female; mean age = 23.45 years, SD = 3.9) participated in the study. Data from the 19 participants (11 male, 8 female; mean age = 22.8 years, SD = 3.6) who completed the study were included in the present analysis. The study was Institutional Review Board (IRB) approved \cite{adhanom2022vr}.

\subsection{Measures}

\subsubsection{Cybersickness Measurements}\label{sec:measures}

Cybersickness was assessed using two complementary measures: repeated subjective discomfort ratings and SSQ scores. 
During immersion, participants provided discomfort ratings on a 0--10 scale once per minute, enabling repeated assessment of symptom progression throughout the VR experience. Unlike prior analyses that aggregated these ratings at the session or trial level, this work preserved the time-resolved discomfort measurements to examine how event-derived eye-movement behavior evolves alongside discomfort progression throughout VR exposure.
SSQ was administered before and after VR exposure to assess overall cybersickness severity across the session. We extracted the SSQ oculomotor subscale \cite{kennedy1993simulator} from the pre- and post-exposure administrations for use in a convergent-validity analysis.

\subsection{Hypotheses}

Prior gaze-cybersickness associations have largely been established through aggregate, session-level, or predictive analyses \cite{nam_eye_2022, ozkan_relationship_2023}, leaving two questions open: whether these associations are observable within shorter temporal windows, and how they evolve as discomfort accumulates over prolonged exposure. We therefore pair a theory-grounded directional hypothesis for each construct with an explicit test of its temporal evolution. We organize our hypotheses according to the three gaze constructs introduced in Section~\ref{sec:metrics}. To model how discomfort-gaze relationships change over time, we treat their temporal evolution as an exploratory question by examining interactions with exposure duration.

\subsubsection{Fixation Stability}

If the perceptual strain of visual-vestibular mismatch taxes gaze stabilization, fixations should become less temporally persistent and less spatially precise as discomfort rises \cite{ebenholtz_motion_1992, foucher2025eye}.

\begin{itemize}
    \item \textbf{H1a:} As discomfort increases, mean fixation duration decreases.
    \item \textbf{H1b:} As discomfort increases, mean fixation dispersion increases.
\end{itemize}

\subsubsection{Visual Exploration}

If discomfort reduces visual exploration, eye movements should become smaller and slower \cite{ozkan_relationship_2023}.

\begin{itemize}
    \item \textbf{H2a:} As discomfort increases, mean saccade amplitude decreases.
    \item \textbf{H2b:} As discomfort increases, mean saccade velocity decreases.
\end{itemize}

\subsubsection{Gaze Allocation}
If discomfort increases central visual attention, gaze should become more center-biased \cite{lee_towards_2025}.
\begin{itemize}
    \item \textbf{H3:} As discomfort increases, mean gaze eccentricity decreases.
\end{itemize}

For each of these hypotheses, we additionally examine whether the relationship changes over exposure duration.

\section{Results}

We used linear mixed-effects models (LMMs) \cite{baayen2008mixed}, fitted with the \texttt{lme4} package in R \cite{bates2015fitting} using restricted maximum likelihood, to examine how discomfort and exposure duration relate to eye gaze behavior. Each gaze metric was modeled as a continuous outcome, with discomfort treated as a continuous predictor despite being ordinal, consistent with common practice in VR research \cite{venkatakrishnan2024effects, deguzman2025reduction}. Fixed effects were discomfort, exposure duration, and their interaction, allowing us to assess both the marginal effect of discomfort and how that effect evolves over time. 

Although eye movements vary with the content of the viewed scene \cite{tatler2011eye}, this variation is largely shared across participants here: all participants navigated the same predefined waypoint route, so the visual scene at a given point in the session was determined chiefly by the route and by elapsed exposure time. The task itself was continuous waypoint-following with free viewing and no discrete stages or stimulus-locked demands, so there are no task phases to model separately. Including exposure duration in the model absorbs this shared temporal and scene-related structure, so the discomfort effect is identified from differences in rated discomfort at matched exposure rather than from where a participant was in the environment.

Discomfort and exposure duration were each centered within participants, so coefficients reflect within-person effects. Participant identity was included as a random intercept to absorb baseline differences across individuals. 
$p$-values were adjusted using the Holm--Bonferroni procedure \cite{holm1979simple} to control the family-wise error rate.

\begin{figure}[htbp]
    \centering
    \includegraphics[width=1.00\linewidth]{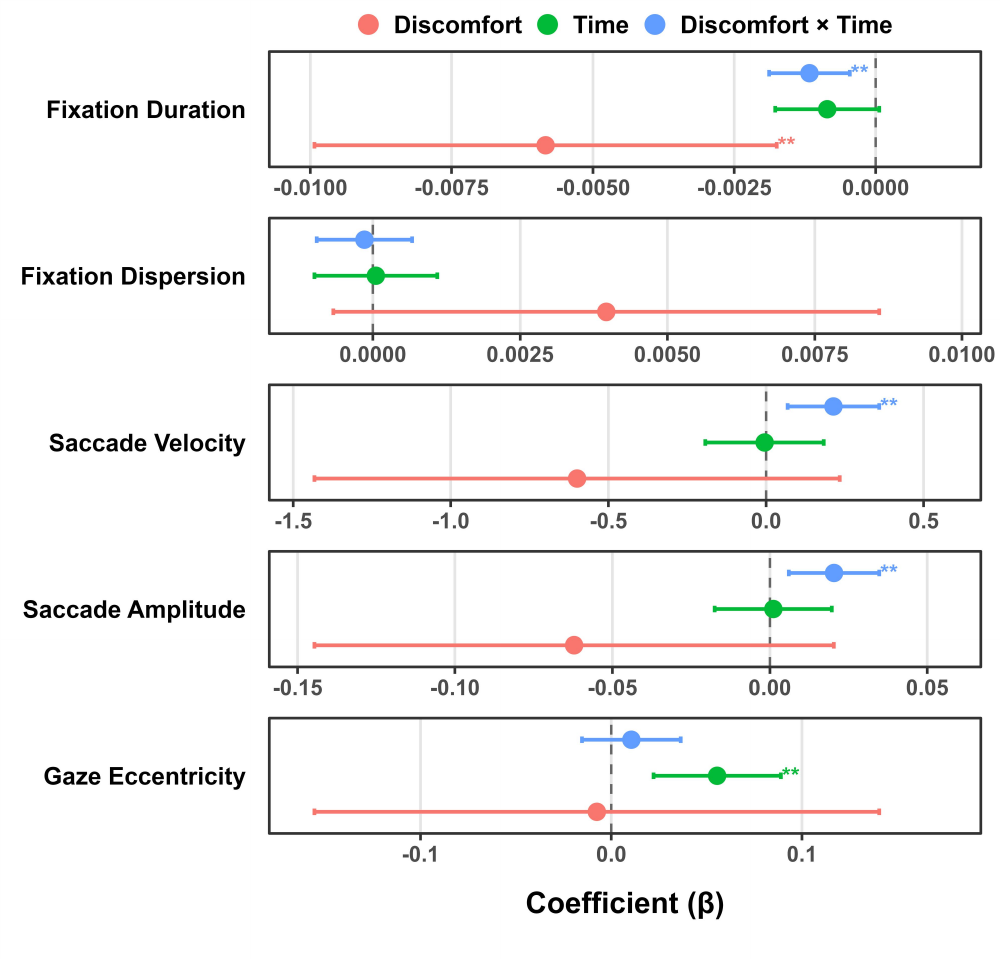}
    \Description{Forest plots showing fixed-effect coefficient estimates and 95 percent confidence intervals for five eye-movement metrics: fixation duration, fixation dispersion, saccade amplitude, saccade velocity, and gaze eccentricity. Coefficients are shown for discomfort, exposure time, and their interaction.}
    \caption{Forest plots of fixed-effect estimates from the interaction LMMs for window-level summaries of event-derived eye-movement metrics. Fixation and saccade events were identified from the eye-tracking signal, and their derived metrics were aggregated within consecutive discomfort rating intervals before modeling. Points represent estimated coefficients, and horizontal bars indicate 95\% confidence intervals. Significance levels are denoted as * $p < .05$, ** $p < .01$, and *** $p < .001$.}
    \label{fig:forest}
\end{figure}


\subsection{Fixation Stability (H1)}
Fixation duration provided the strongest evidence in the study. Higher discomfort was associated with significantly shorter fixations ($\beta = -0.0058$, $SE = 0.0021$, $\beta_{\text{std}} = -0.11$, $p = .005$, $p_{\text{Holm}} = .027$), \emph{supporting H1a}, and this association strengthened over exposure: the discomfort $\times$ exposure interaction was significant and negative ($\beta = -0.0012$, $SE = 0.0004$, $\beta_{\text{std}} = -0.12$, $p = .001$, $p_{\text{Holm}} = .007$). Discomfort thus shortened fixations on average, and increasingly so as the session progressed.

\begin{figure*}[t]
  \centering
  \includegraphics[width=\textwidth]{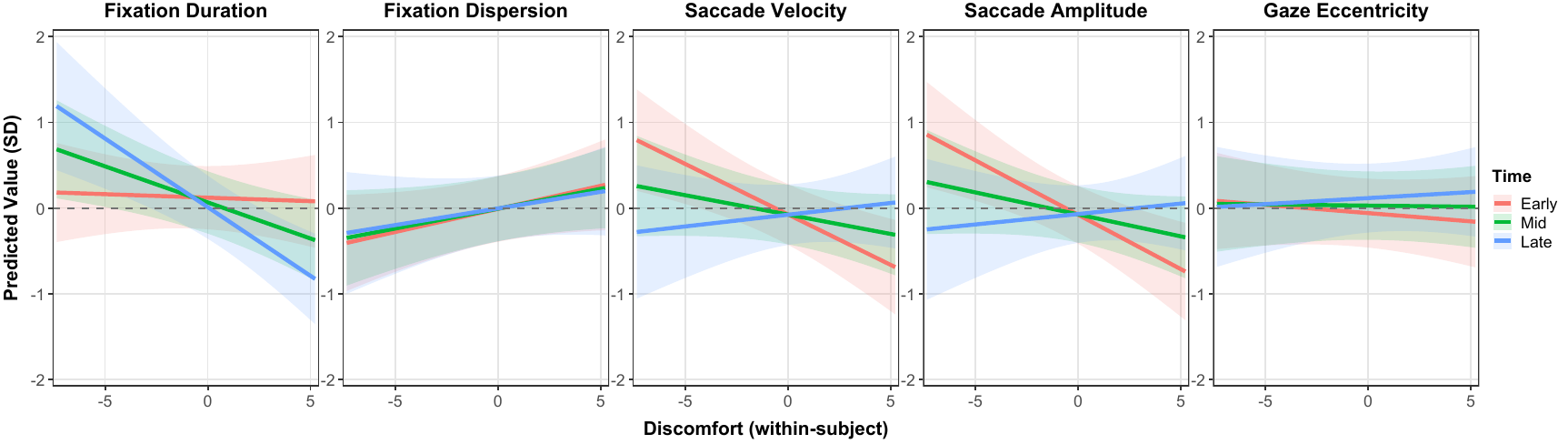}
    \caption{Predicted eye-movement metrics as a function of within-subject centered discomfort at three representative time points (early, mid, and late exposure; 25th, 50th, and 75th percentiles of centered exposure time). The negative relationship between discomfort and fixation duration becomes steeper over time, while saccade velocity and saccade amplitude exhibit time-dependent changes in their relationships with discomfort. Fixation dispersion shows no association with discomfort and no interaction with time. Gaze eccentricity increases significantly with exposure time but shows no association with discomfort and no discomfort-by-time interaction. Lines show standardized model predictions (in SD units), and shaded areas represent 95\% confidence intervals.}
    \Description{Five side-by-side interaction plots for fixation duration, fixation dispersion, saccade amplitude, saccade velocity, and gaze eccentricity. Each panel shows standardized predicted values across within-subject centered discomfort, with separate prediction lines and confidence bands representing early, mid, and late exposure. Fixation duration shows increasingly negative slopes across the three exposure levels. Saccade amplitude and saccade velocity also show differences in their relationships with discomfort across exposure levels. Fixation dispersion and gaze eccentricity show relatively little change with discomfort.}
  \label{fig:fixation_duration}
\end{figure*}

In contrast, fixation dispersion showed neither a discomfort effect ($\beta = 0.0040$, $SE = 0.0024$, $\beta_{\text{std}} = 0.06$, $p = .095$, $p_{\text{Holm}} = .38$) nor a change over exposure (interaction $p_{\text{Holm}} = .85$). \emph{H1b was not supported.} Cybersickness therefore compressed the temporal persistence of fixations, progressively so, without measurably broadening their spatial spread.

\subsection{Visual Exploration (H2)}
Neither saccadic measure showed a reliable marginal effect of discomfort, but both showed a significant change in their relationship to discomfort over exposure. For saccade amplitude, the discomfort effect was not significant ($\beta = -0.062$, $SE = 0.042$, $\beta_{\text{std}} = -0.07$, $p = .14$, $p_{\text{Holm}} = .42$), so \emph{H2a was not supported}; however, the discomfort $\times$ exposure interaction was significant and positive ($\beta = 0.020$, $SE = 0.0073$, $\beta_{\text{std}} = 0.12$, $p = .006$, $p_{\text{Holm}} = .017$). Saccade velocity mirrored this pattern: the main effect of discomfort was not significant ($\beta = -0.60$, $SE = 0.43$, $\beta_{\text{std}} = -0.06$, $p = .16$, $p_{\text{Holm}} = .42$), leaving \emph{H2b unsupported}. However, the discomfort $\times$ exposure interaction was significant and positive ($\beta = 0.214$, $SE = 0.074$, $\beta_{\text{std}} = 0.12$, $p = .004$, $p_{\text{Holm}} = .017$).

In both cases, the early negative association between discomfort and saccadic exploration weakened and tended to reverse as exposure continued. The two measures showed near-identical interaction magnitudes ($\beta_{\text{std}} = 0.12$ for both), suggesting a coordinated, time-varying adjustment of visual exploration rather than an isolated change in a single saccadic parameter.

\subsection{Gaze Allocation (H3)}
Gaze eccentricity showed no association with discomfort ($\beta = -0.0076$, $SE = 0.076$, $\beta_{\text{std}} \approx 0.00$, $p = .92$, $p_{\text{Holm}} = .92$) and no change in that relationship over exposure (interaction $p_{\text{Holm}} = .85$). \emph{H3 was not supported.} Unlike the center-biased viewing reported for 360$^\circ$ video, discomfort did not concentrate gaze toward the forward direction during interactive locomotion. Eccentricity did, however, increase with exposure duration independent of discomfort ($\beta = 0.056$, $SE = 0.017$, $p = .001$), indicating a time-driven rather than discomfort-driven shift in gaze allocation.

\subsection{Convergence with SSQ Oculomotor Symptoms}

As an exploratory convergence check, we related each participant's post-exposure SSQ oculomotor subscale score to their mean value of each gaze metric using Spearman correlations, Holm--Bonferroni corrected across the five metrics. No metric was associated with oculomotor symptom scores. For fixation stability, neither fixation duration ($\rho = .04$, $p = .88$, $p_{\mathrm{Holm}} = 1.00$) nor fixation dispersion ($\rho = .09$, $p = .73$, $p_{\mathrm{Holm}} = 1.00$) was related to subjective symptoms. For visual exploration, neither saccade amplitude ($\rho = -.07$, $p = .77$, $p_{\mathrm{Holm}} = 1.00$) nor saccade velocity ($\rho = -.03$, $p = .91$, $p_{\mathrm{Holm}} = 1.00$) showed a reliable association. Gaze allocation, indexed by gaze eccentricity, likewise showed none ($\rho = .37$, $p = .12$, $p_{\mathrm{Holm}} = .60$). Because the SSQ is administered once after exposure, it indexes overall session severity and does not resolve the within-session dynamics that are the focus of this analysis. We therefore do not read these null results as evidence against the within-session effects, which are estimated across event-derived windows with greater resolution. They instead suggest that oculomotor discomfort relates to gaze through its within-session progression rather than as a static, between-person trait.

\section{Discussion}

This study investigated how eye-movement behavior evolves during the temporal progression of cybersickness in immersive VR. Overall, the results indicate that fixation stability is directly associated with increasing discomfort, while the relationship between saccadic behavior and discomfort changes over the course of VR exposure. In addition, gaze allocation remained largely stable with respect to discomfort. These findings extend prior work by moving beyond aggregated or predictive analyses of gaze data to characterize how oculomotor behavior dynamically evolves during prolonged VR exposure.

\subsection{Fixation Stability}
Fixation duration provided the clearest signal. Higher discomfort was associated with shorter fixations, supporting H1a, and the negative discomfort\,$\times$\,exposure interaction indicates that this association strengthened as the session progressed. The substantive point is not simply that fixations shorten under discomfort, which prior work anticipates \cite{nam_eye_2022, rayner2009eye, foucher2025eye}, but that discomfort's influence on gaze stabilization deepened with continued exposure rather than acting as a fixed perturbation. This is consistent with a stabilization system whose tolerance to sensory conflict erodes cumulatively, such that sustained conflict exacts a progressively larger behavioral cost. Fixation dispersion, in contrast, showed neither a discomfort association nor a change over exposure, providing no support for H1b. This dissociation is itself informative: cybersickness did not broadly destabilize fixation but specifically compressed its temporal persistence while leaving its spatial precision intact, arguing against a generalized ``gaze becomes noisier'' account \cite{nam_eye_2022} and identifying fixation duration as the more diagnostic of the two facets. This finding also complements Tovar et al.'s \cite{tovar_identifying_2024} finding that blink behavior most strongly tracked discomfort, together suggesting that which oculomotor channel signals discomfort may depend on the source of the discomfort.

\subsection{Visual Exploration}

Saccadic behavior showed a different pattern. Neither saccade amplitude nor saccade velocity exhibited a reliable marginal association with discomfort, so H2a and H2b were not supported; both, however, showed significant and
near-identical positive discomfort\,$\times$\,exposure interactions ($\beta_{\text{std}} = 0.12$ for both). Rather than indicating that saccadic behavior is simply unrelated to discomfort, this pattern suggests that the
relationship is not stable across the session. We refrain from characterizing the early-exposure end of this interaction as an established reduction in saccadic exploration because the marginal effect was not
significant.
Thus, the interaction demonstrates only that the relationship between discomfort and saccadic behavior changes over time, not its sign or significance at any particular stage of exposure.
This implies that visual exploration does not respond uniformly to discomfort but is reorganized as exposure continues, consistent with a strategy-level adjustment in how the scene is sampled rather than a fixed oculomotor consequence of discomfort.

This reorganization seems to conflict with the strengthening fixation effect, but the two may fit together rather than clash. One possible explanation may be that gaze behaviors users have little control over are progressively disrupted by sustained conflict, while behaviors they can control are adjusted as discomfort rises, shorter fixations reflecting the first, the shifting 
saccadic
relationships the second. We offer this as a possible reading, not a confirmed mechanism, since the distinction is not clear-cut and our metrics do not separate the underlying control processes.

\subsection{Gaze Distribution}

\begin{figure}[htbp]
  \centering
  \includegraphics[width=1.00\linewidth]{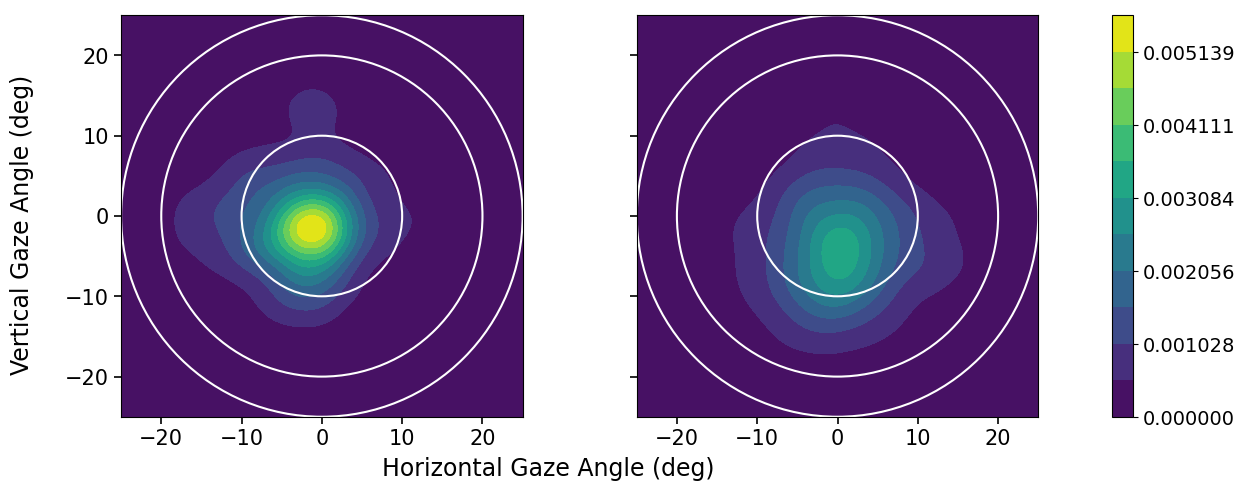}
  
   \caption{Spatial gaze density heatmaps aggregated across participants for the first and last temporal decile (10\%) of the VR session.
   The left panel shows the early phase, and the right panel shows the late phase of VR exposure.
   Fixation density is more concentrated during the first minutes of exposure, whereas later exposure shows a more dispersed gaze distribution.}
   \Description{Two side-by-side spatial gaze density heatmaps plotted by horizontal and vertical gaze angle. The early-exposure heatmap on the left shows a strong, compact density peak near the center of the visual field. The late-exposure heatmap on the right shows a weaker and more broadly distributed central region, indicating greater spatial dispersion of gaze. Concentric reference circles surround the center in both panels.}
  \label{fig:gaze_heatmap_10}
\end{figure}

Gaze allocation showed a third pattern. Gaze eccentricity was associated with neither discomfort nor the discomfort $\times$ exposure duration interaction, providing no support for H3. However, it increased reliably with exposure duration independent of discomfort. This finding indicates that visual attention gradually shifted farther from the central visual field as the session progressed (see Figure~\ref{fig:gaze_heatmap_10}). Although this increase might initially suggest broader visual exploration, when considered alongside progressively shorter fixations and the evolving pattern of saccadic behavior, it may instead reflect a gradual decline in organized visual scanning. Rather than indicating more effective exploration of the environment, the gradual increase in gaze eccentricity may reflect a reduced ability to maintain stable visual attention as the effects of prolonged exposure accumulate, resulting in gaze being directed more frequently toward peripheral regions. We present this as a plausible hypothesis rather than a confirmed mechanism.

Taken together, these findings suggest that cybersickness does not uniformly affect the oculomotor system. Instead, different components of gaze exhibit different temporal responses as discomfort develops, indicating that eye-movement behavior undergoes a progressive reorganization rather than a single, global disruption during prolonged VR exposure. More broadly, this work demonstrates the value of examining event-derived gaze metrics in conjunction with repeated measures of discomfort, revealing temporal dynamics that would likely remain obscured by aggregate or session-level analyses.
By characterizing how different aspects of gaze evolve over time, this study provides a more nuanced understanding of the behavioral progression of cybersickness.
From a study-design perspective, the findings of this study suggest that future cybersickness research may benefit from combining repeated symptom assessments with temporally resolved gaze measures and accounting for exposure duration, as session-level aggregation may obscure within-participant changes and evolving relationships between oculomotor behavior and cybersickness over the course of VR exposure.

\subsection{Limitations and Future Work}

The findings should be interpreted in light of several limitations. First, the analysis was conducted on a modest sample of 19 participants within a single VR environment, navigation task, and locomotion paradigm, which may limit statistical power to detect smaller effects and the generalizability of the findings to other VR experiences. Future work using larger and more diverse samples, multiple environments, and different locomotion methods could establish the robustness of the observed relationships across VR experiences.

A second limitation is that cybersickness-specific changes cannot be fully separated from other processes that unfold over prolonged exposure. Because discomfort accumulates together with elapsed time within each session, the discomfort and exposure-duration effects are inherently correlated, and this observational design cannot fully distinguish cybersickness-specific changes from general time-on-task processes such as fatigue or adaptation. Because cybersickness develops naturally during navigation rather than being independently manipulated, an otherwise equivalent no-cybersickness control condition is nontrivial. Designs that vary cybersickness-related stimulation independently of exposure duration could more directly separate these influences.

Finally, eye-tracking measurements during prolonged VR exposure may be affected by measurement error, calibration drift, or headset slippage over time. Because intermediate recalibration measurements were not available, these factors cannot be fully excluded as alternative contributors to the observed oculomotor patterns. Such drift would be expected to degrade the spatial metrics, fixation dispersion and gaze eccentricity, most; yet neither was associated with discomfort, whereas the discomfort effect was confined to fixation duration, a temporal measure comparatively insensitive to a gradual
spatial offset. This dissociation makes tracking error an unlikely explanation for the central result, although periodic calibration checks in future studies would allow it to be tested directly.

Beyond these experimental considerations, future work could investigate whether nonlinear models capture more complex relationships between discomfort, exposure duration, and oculomotor behavior, as well as explore the use of these temporal gaze patterns for real-time cybersickness prediction and adaptive VR systems.

\section{Conclusion}

This study investigated how eye-movement behavior evolves during the temporal progression of cybersickness in immersive VR exposure. The results show that fixation duration decreases with increasing discomfort, with the relationship strengthening over time, whereas saccade amplitude and saccade velocity were not directly associated with discomfort but exhibited significant time-dependent relationships with discomfort. In contrast, fixation dispersion remained stable, and gaze eccentricity was influenced by exposure duration rather than discomfort. Together, these findings demonstrate that the relationship between eye-movement behavior and cybersickness is dynamic, with different oculomotor metrics exhibiting distinct temporal responses during prolonged VR exposure. By focusing on interpretable event-derived eye gaze metrics alongside repeated discomfort measurements, this work provides a behavioral characterization of how oculomotor dynamics evolve throughout cybersickness progression. Overall, these findings highlight the importance of incorporating temporal dynamics into eye-tracking analyses to better understand how oculomotor behavior evolves during cybersickness in immersive VR.

\bibliographystyle{ACM-Reference-Format}
\bibliography{references}

\end{document}